\documentclass[aps,prb,twocolumn,superscriptaddress,showpacs,floatfix]{revtex4-1}

\usepackage{graphicx}
\usepackage{color}
\begin{document}
%
\preprint{}
%
\title{Experimental determination of tunneling characteristics and dwell
times from temperature dependence of Al/Al$_2$O$_3$/Al junctions}
\author{Edgar J. Pati\~no and N. G. Kelkar}
\affiliation{Departamento de F{\'i}sica, Universidad de los Andes,
Cra.1E No.18A-10, Bogot\'a, Colombia}
%
%
\date{\today}
%
%
\newcommand{\Ce}{C$_E\,$}
\newcommand{\Wm}{\omega_\times}
\begin{abstract}
Measurements of current-voltage (I-V) characteristics of a high quality
Al/Al$_2$O$_3$/Al junction at temperatures ranging from 3.5 K to 300 K have been
used to extract the barrier properties. Fitting results
using Simmons' model led to a constant value of barrier width
$s$$\sim$20.8~$\textrm{\AA}$ and a continuous increase in the barrier height with decreasing temperature.
The latter is used to determine the energy band gap temperature dependence and average phonon frequency
$\omega$ = 2.05 $\times$ 10$^{13}$ sec$^{-1}$ in Al$_2$O$_3$, which adds confidence to the
precision of our measurements. The barrier parameters are used to
extract the temperature dependent dwell
times in tunneling ($\tau_D$ = 3.6 $\times$ 10$^{-16}$ sec at mid-barrier energies)
and locate resonances above the barrier.
\end{abstract}
%
\maketitle



Quantum tunneling has been widely used
in order to investigate
the density of states of the materials across the barrier \cite{kontos,boden}
and magnetoresistance in magnetic tunnel junctions (MTJs).
In most tunnel junction experiments ultra thin Al$_2$O$_3$  has been
the standard choice for barrier material.
In spite of the possible applications
there is no clear understanding of the barrier parameters as a function of temperature.
Experiments carried out previously have found an ``anomalous" temperature
dependence of the current voltage characteristics of Al$_2$O$_3$ tunnel
junctions \cite{gundlach,nelson,kadlec,das,meyerhofer}.
The characteristics were originally termed anomalous since the temperature
dependence was found to be
greater than that corresponding to the temperature change of the Fermi-Dirac distribution of electrons in the electrodes.
Using Simmons model \cite{simmons1,simmons2},  it was found that the metal-insulator barrier height decreases with increasing temperature.
The possible explanations have been controversial. This effect has been attributed to the band gap temperature dependence, the change in the dielectric space charge and trap levels in the insulator \cite{meyerhofer,das}. This last mechanism, if true, would prevent the electrons from direct tunneling \cite{bahlouli}.
On the other hand, the barrier width is a parameter which has been difficult to interpret given the fact that it
need not be the same as the oxide thickness as it also depends on image forces  \cite{simmons1,simmons2}.
There are two reports using Simmons's model on variations of the barrier width as temperature changes \cite{das,akerman}.  As a result of barrier shorts in MTJs
\cite{akerman}, the apparent barrier parameters extracted from Simmons's model showed
an increase of barrier width with decreasing temperature. Conversely, a small decrease of
barrier width was also reported in \cite{das, akerman} without any clear explanation.
For ultrathin Al$_2$O$_3$ barriers other complications such as pinholes, barrier shorts and barrier regions thinner than the average barrier thickness (i.e. ``hot spots") may arise in the fabrication. These may even occur in experiments where advanced plasma oxidation techniques have been used \cite{akerman, dorneles}.

The present work focuses on the temperature dependence of the tunneling phenomena across Al$_2$O$_3$.
We find no appreciable variation of the barrier width as a function of
temperature and a mild temperature dependence of the barrier height.
Furthermore, using barrier height temperature dependence we
demonstrate that it is possible to directly link the barrier height and the energy gap
of ultra thin Al$_2$O$_3$, assuming a simple linear relation. Thus we are able to
extract important quantities,
not well known, such as band gap at zero temperature, average phonon frequency and the coupling constant
related to electron phonon coupling \cite{chiu}.
The phonon frequency when compared with the value reported in literature from
sophisticated experiments corroborates the accuracy of our experimental findings.
These results lead us to conclude that the main contribution to the temperature dependence of the I-V characteristics in Al$_2$O$_3$ tunnel junctions comes
from the semiconductor band gap temperature dependence
leaving out dielectric space charge effect and trap levels in the insulator.
Finally we used the experimentally extracted
barrier height and width parameters to calculate the tunneling time,
which, is being reported here for a solid state tunnel junction.
The order of magnitude of this time corresponds to the one obtained in sophisticated experiments involving laser
induced tunneling of electrons \cite{NatSci}.

\begin{figure}[h]
\includegraphics[width=8.5cm,height=13.4cm]{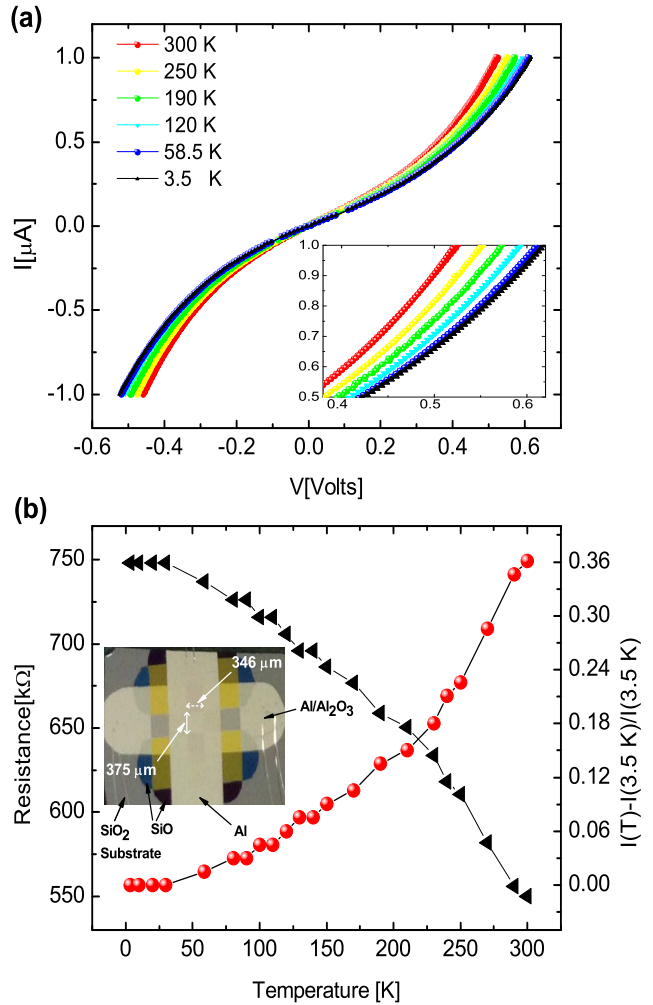}
\caption{\label{fig:fig1}
(a) I-V characteristics of Al/Al$_2$O$_3$/Al junctions at different temperatures;
inset shows zoom in view at upper voltages. (b) Junction Resistance (triangles) and Tunneling current variation (circles) vs
temperature at a bias voltage of 0.5 V; inset shows top view
of Al/Al$_2$O$_3$/Al junction. The junction area is defined by SiO edges between top and bottom Al electrodes.}
\end{figure}

High quality Al/Al$_2$O$_3$/Al planar junctions were fabricated
as a result of carefully proceeding with the following steps.
Given the high Al$_2$O$_3$ resistivity, Aluminium junctions on Silicon
require a very good electrical insulation from the Si layer. Therefore,
a thick SiO$_2$ layer is needed as a base layer in order to prevent current
runaway through the substrate. For this reason the Si(100) substrate was oxidized
under 1100$^{o}$ C in an oven for a 6 hour period.
This permitted to obtain a thick SiO$_{2}$ layer of $\sim$300 nm.
Using a HV e-gun evaporation system, each of the layers were grown in a
chamber with a base pressure better than 1$\times$10$^{-7}$ Torr and evaporation
pressure less than 1$\times$10$^{-6}$ Torr.
The film thickness was monitored during growth to better than
1 $\textrm{\AA}$ by quartz balance.
Thin Aluminium layers ($\sim$30 nm) grown in this way were very uniform and
did not show any pinholes as verified by optical imaging.
Our Al/Al$_2$O$_3$/Al junctions were fabricated using a series of mechanical masks
(inset Fig.~\ref{fig:fig1}b).
The first step was to evaporate a 30 nm Al bottom electrode under a high deposition
rate of about a 10 $\textrm{\AA}$/sec.
Then the sample was removed from the chamber
and allowed to oxidize in air at room temperature for a period longer than 20 days.
This oxidation method was preferred given its lowest oxidation rate as
compared to more sophisticated methods such as oxygen plasma and dry oxygen methods.
The former may lead to shorts and pinholes or ``hot spots" \cite{akerman, dorneles} and the latter compared to
the present work does not produce a more robust oxide layer as will be discussed later.

In order to define the junction geometry,
50 nm of SiO was evaporated in two steps through masks after oxidation
totaling to 100 nm of insulation.
Using a final mask, the top Al electrode ($\sim$30 nm) was deposited.
Tunnel junction areas of 375 $\mu$m~$\times$~346 $\mu$m were obtained as checked by
optical microscope measurements (Inset Fig.~\ref{fig:fig1}b).
The above fabrication process was the result of preliminary experiments on more than 20 junctions of different sizes and oxidation times leading to the final device presented here. The present results can be reproduced by carefully repeating the above steps.
\\
Current-Voltage (I-V) characteristics were acquired in a four terminal
geometry.
Here a high precision dc current source provided the current while the voltage was
measured using a nano-voltmeter. A large number of I-V characteristics were
obtained in the temperature range between 3.5 K and 300 K.
Prior to each measurement, the temperature stability was better than 1 mK.
The current and voltage through the junction were kept below 1 $\mu$A and
0.7 V respectively in order to prevent overheating.
Sufficient heat sinking was demonstrated in the junction as no hysteresis was
observed after cycling the current two or more times at each temperature.
Furthermore, performing measurements with the device on different days
demonstrated high reproducibility of the I-V characteristics over time.
The results are shown in Fig.~\ref{fig:fig1}a where only a few
I-V curves are depicted out of a total of 23.
To check the barrier quality, the resistance vs temperature (RT)
characteristic of the tunnel junction was obtained, at 0.5 V bias voltage, indicating an insulator like
temperature dependence of R  (triangles Fig.~\ref{fig:fig1}b). More importantly a junction area resistance product $A$$R_{N}$$\sim$6$\times$10$^{10}$$\mu$m$^{2}$ was calculated at high voltages. This value is substantially higher than those of other works using pure oxygen \cite{boden} and specialized oxidation techniques \cite{kontos, sangiorgio} confirming the superior quality of our junctions.
The tunneling current variation as a function of temperature was also obtained at a bias voltage of 0.5~V (Fig.~\ref{fig:fig1}b circles). Here an increase of about 36$\%$ is observed between 3.5 and 300 K.
Clear differences of the IV curves and tunneling currents demonstrate the ``anomalous" temperature dependence of the current voltage characteristics.
Note also that I-V characteristics are slightly asymmetrical for voltages larger
than $\sim$0.1 V depending on the electrode's polarity.
When the top electrode is positively charged (forward bias) the tunneling
current is slightly lower as compared to the negatively charged polarity (back bias).
This is the result of an effective barrier height which is slightly higher in
the forward bias direction and lower in the reverse bias direction of the current.
In order to investigate this further, the I-V data was carefully fitted using Simmons's
model \cite{simmons1,simmons2} for the case of a symmetric rectangular barrier
and a free electron effective mass. All data analyses
have been carried out using forward and backward current
direction. Given that the results show only very small
variations depending on current direction we focus on the forward bias experiment.

\begin{figure}[h]
\includegraphics[width=8.5cm,height=6.0cm]{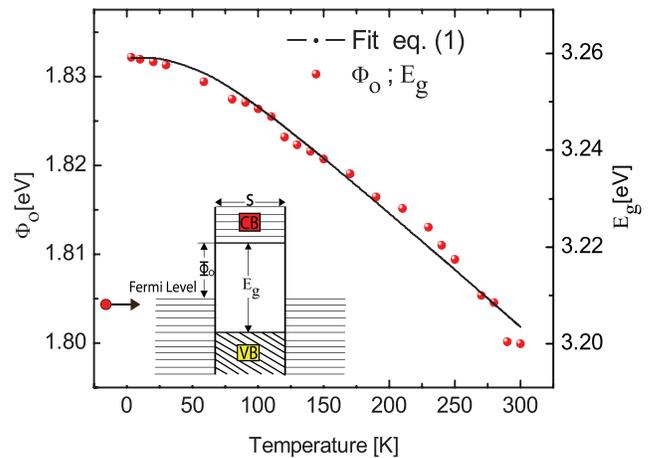}
\caption{\label{fig:fig2}
Barrier height and energy gap as a function of temperature as extracted from
Simmons's model and deduced from $E_g$(T)=$\gamma$ $\Phi_0$(T) respectively.
Inset: Schematic electron energy diagram for a Metal-Semiconductor-Metal
junction without voltage bias. The solid line displays the fit for the energy gap using
Eq. (\ref{energygap}).}
\end{figure}

The best fit to our data was obtained when image forces were not included.
This is due to the fact that in materials with large dielectric constant K$\ge$4
the contour of a practical barrier follows closely that of a
rectangular barrier \cite{simmons1,simmons2}.
For a dielectric material such as Al$_2$O$_3$ with K$\ge$8 (see \cite{carmona}
and references therein) this is clearly the case.
The junction parameters; barrier height ($\Phi_{0}$), barrier width ($s$)
and junction area ($A$) were determined in the intermediate voltage regime, i.e.,
0$\ge$V$\le$$\Phi_0$; using the following procedure at each temperature.
First a three parameter fit is performed. Remarkably, the junction area thus obtained corresponds to
the value measured by the optical microscope. As explained in reference \cite{dorneles}
this corroborates the absence of ``hot spots" and pinholes in our junction.
In the next step we fixed $A$ and one of the two parameters; $\Phi_{0}$ or $s$,
while the remaining other one was extracted from an effective one parameter fit.
This procedure was repeated self consistently.
This led to a constant value of barrier width
$s$$\sim$20.8~$\textrm{\AA}$
with a standard error less than 0.002 at all temperatures \cite{Note}.
On the other hand, $\Phi_0$ was found to continuously decrease with temperature
between the values of
1.799 eV at 300 K and 1.83 eV at 3.5 K.
The barrier height thus increased as the temperature decreased from 300 to 3.5 K
with a $\Delta\Phi_0$~$\sim$~33~meV as depicted in Fig.~\ref{fig:fig2}.
The values found for the barrier width, height and variation are very close to
the ones found in previous works
(see \cite{gundlach,kadlec,das} and references therein).
Further analysis of our data can be obtained considering the illustration shown
in inset Fig ~\ref{fig:fig2}  where the energy gap ($E_g$) and barrier height
($\Phi_0$) can be directly related in the following way.
The barrier height in a tunnel junction depends on the semiconductor energy gap.
Assuming a linear relation between these two parameters, we take the
ratio $E_g$(T)/$\Phi_0$(T) = $\gamma$ to be constant at all temperatures.
Therefore we may calculate $\gamma$ = $E_g$(300K)/$\Phi_0$(300K) from room
temperature measurements. Here Eg(300K)=3.2~eV obtained from \cite{costina} for amorphous Al$_2$O$_3$ and $\Phi_0$(300K)=1.799~eV from our current data fit.
Finally we obtain the temperature dependence of the energy gap from
$E_g$(T)=$\gamma$ $\Phi_0$(T) in the temperature range between
3.5 and 300 K as given in Fig.~\ref{fig:fig2}.

Using the temperature dependence of the semiconductor energy gap data we
can obtain additional information by fitting our experimental results to the
following theoretical expression;
\begin{equation}\label{energygap}
E_g(T) = {E_g(0)-S\langle\hbar \omega\rangle[\text{coth}(\langle\hbar \omega \rangle/2kT)-1] } \, ,
\end{equation}
proposed by O'Donnel and Chen \cite{odonnell} as a replacement of the more commonly used
Varshni equation. The authors find the formula compatible with assumptions regarding
the influence of phonons on the energy band gap.
$E_g(0)$ in (\ref{energygap}) is the band gap at zero temperature,
$S$ a dimensionless coupling constant and $\langle\hbar \omega\rangle$
is the average phonon energy. We obtain an excellent fit with the values \cite{comment}
$E_g$(0)=3.26 eV (standard error 0.001), S = 1.414 (standard error 0.065) and
$\langle\hbar \omega\rangle$=13.5~meV (standard error 0.0016) as
can be seen in Fig ~\ref{fig:fig2} solid line. This fit allows us to determine an
average phonon frequency of $\langle\omega\rangle$ = 2.05$\times$10$^{13}$ sec$^{-1}$.
In order to compare this value with other experiments, we use the speed of
sound measurements of amorphous Al$_2$O$_3$ ($v_{Al_{2}O_{3}}$) obtained using the
picosecond ultrasonic technique by Rossignol et. al.\cite{rossignol}.
Here speed of sound yields the longitudinal sound velocity $v_{Al_{2}O_{3}}$
= 6.7$\times$10$^{3}$ m/s. Considering a lattice parameter
$a\sim4.7$ $\textrm{\AA}$ and a value of k=$\pi$/2$a$ at the middle of the
first Brillouin zone, from the expression $\omega$ = $v_{Al_{2}O_{3}}$ k;
the phonon frequency $\omega$ = 2.24 $\times$ 10$^{13}$ sec$^{-1}$ is obtained.
Indeed these two values are exceedingly close giving us confidence on the accuracy of
our measurements.
\begin{figure}[h]
\includegraphics[width=8cm,height=10cm]{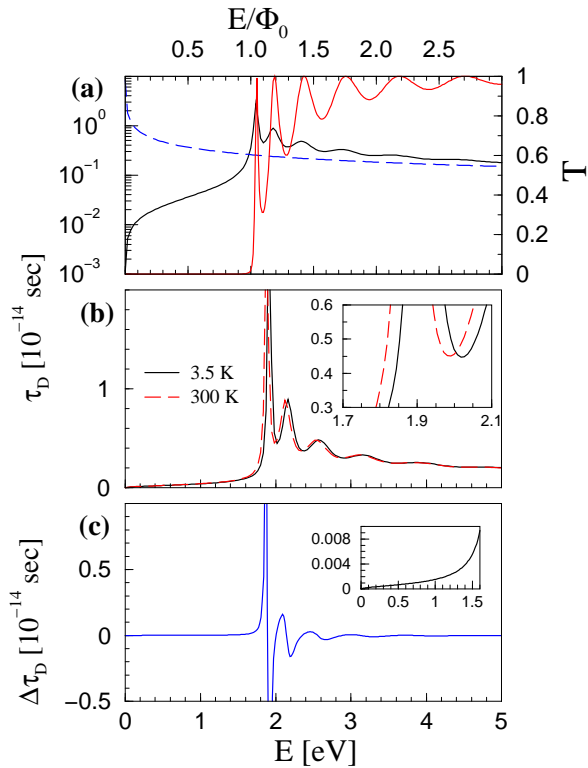}
\caption{\label{fig:fig3}
Average dwell times within the potential barrier shown (a) as a function of the
energy of the tunneling electrons divided by the barrier height and (b)
as a function of the energy for two different temperatures.
Though the difference between the dwell time curves at 3.5 and 300 K in
general is negligible, in (c) we see that it becomes pronounced mostly in the
resonance regions ($\Delta \tau_D$ = $\tau_D^{300 K} - \tau_D^{3.5 K}$).
The transmission coefficient is given by the red line with scale on the
right side in (a).}
\end{figure}
\\
Having extracted the tunneling barrier heights from the I-V characteristics
data, we now turn to the evaluation of the time spent by the electrons in the
barrier region. Though the time spent by the electron interacting with the
semiconductor while tunneling is expected to be tiny and not directly
measurable, a knowledge of the barrier parameters obtained from data does
enable us to calculate this time. We choose to calculate the so-called
average dwell time of the electron in the barrier from among the
several definitions available in literature. This time concept was first
introduced by Smith \cite{time2} in the context of quantum collisions
as the average time spent by the interacting particles in a given region of
space and seems to have emerged over the years
as one of the most useful concepts with a physical
significance \cite{time3}. For an arbitrary barrier $V(x)$ in one-dimension,
the standard dwell time for a particle with energy $E$ is given as \cite{buett}
$\tau_D (E) = [\int_{x_1}^{x_2} \, |\Psi (x)|^2 \, dx]/j$,
where $x_1$, $x_2$, are the classical turning points corresponding to
$E = V(x)$.
Intuitively, the time spent by the particle with velocity $v$
can be written as $t = \int_{x_1}^{x_2} \,dx / v = \int \, \rho(x) \,dx /j(x)$,
where the probability current density $j = \rho v$ with the probability density
$\rho = |\Psi(x)|^2$ for $\Psi(x)$ which is a solution of the time independent
Schr\"odinger equation.
Since $\rho$ is time independent, the continuity equation
$\vec{\nabla} \cdot \vec{j} = 0$ and $j$ ($= j_x$ here) is a constant.
For a plane wave $e^{i k x}$ incident at the barrier,
$j = \hbar k /m$ with $k = \sqrt{ 2 m E}/\hbar$ and we obtain the dwell time
expression given above.

For a rectangular barrier of fixed height and width, $\tau_D(E)$ can be evaluated
analytically and is given for $E < \Phi_0$ as \cite{buett}:
\begin{equation}
\tau_D = {m k \over \hbar \kappa} {2 \kappa s (\kappa^2 - k^2) + k_0^2
\sinh{(2\kappa s)} \over 4 k^2 \kappa^2 + k_0^4 \sinh^2{(\kappa s)}}
\end{equation}
where $k_0 = \sqrt{ 2 m \Phi_0}/\hbar$ and $\kappa = \sqrt{k_0^2 - k^2}$. For
$E > \Phi_0$, one has to replace $\kappa$ by $i K$ where
$K = \sqrt{2 m (E - \Phi_0)}/\hbar$.
Fig. 3 shows the average dwell time (which refers to the time spent in the
barrier regardless of whether the particle was reflected or transmitted)
as a function of the energy of the tunneling electrons at 3.5 K and 300 K.
Fig 3a shows the curves as a function of the energy normalized by the barrier
height (which is different at 3.5 K and 300 K) and Fig 3b shows the dwell times
as a function of the energy in eV.
The time shown in Fig. 3 is essentially the time spent by the electron
in the Al/Al$_2$O$_3$/Al sandwich in the forbidden (tunneling) region.
It is 3.6 $\times$ 10$^{-16}$ sec at mid-barrier energies (Fig. 3a)
and hence smaller than the scale mentioned in \cite{time1} for image forces
to become important.
Resonances start appearing at energies above the barrier but quickly vanish
around $E/\Phi_0$ = 3, where one can see that the time starts
approaching the value for
a free particle traversing the width of the barrier (dashed blue line in Fig. 3a).
It is interesting to note that though the resonance positions in the dwell
time plot are in general the same as those in the
transmission coefficient (T) (red line Fig. 3a), T displays more
prominently the resonances than there actually are in the dwell time curves.
The average dwell time in
tunneling has been shown to have the significance of a density of states
\cite{time5} whereas the peaks in T simply correspond to poles (where
it appears like every pole may not necessarily correspond to a sharp resonance).
In Fig. 3c we plot the difference between
$\tau_D$ at 300 K and 3.5 K and note that it is
pronounced in the region where resonances occur.

Finally, we must mention that though all calculations in Fig. 3 are presented
for the free electron mass, introducing an effective electron mass
(such as 0.75 $m$ as given in \cite{rippard}) changes the barrier height and
width (though only slightly) in such a way that it does not
change the dwell time for electrons with $E < \Phi_0$. For energies above the
barrier,
the dwell time reduces for a smaller effective mass and shifts the resonances
slightly. This is understandable since transmission becomes a classical effect
above the barrier and lighter electrons move faster.

To summarize, we can say that the data on the I-V characteristics
in an Al/Al$_2$O$_3$/Al junction at different temperatures
have been analysed to extract various features in the tunneling of
electrons through this junction. Fitting results using Simmons's model led to a constant value of the barrier width
$s$$\sim$20.8~$\textrm{\AA}$ at all temperatures (contrary to \cite {das}) and a
continuous increase in the barrier height with decreasing temperature.
The direct measurements of quantities such as the phonon frequencies and quantum tunneling times
\cite{NatSci} require sophisticated techniques. The present
work provides an indirect measurement of the same from a rather simple
tunneling experiment.
The average phonon frequency $\langle\omega\rangle$ = 2.05$\times$10$^{13}$ sec$^{-1}$ extracted
from our tunneling experiments is exceedingly close to the one obtained from speed
of sound measurements $\omega$ = 2.24 $\times$ 10$^{13}$ sec$^{-1}$.
It is also remarkable that the order of magnitude of
the tunneling times of the present work which ranges from 10$^{-16}$ sec at mid barrier
energies to 10$^{-14}$ sec in the resonance region is similar to the times
obtained in \cite{NatSci}.
Similar analyses of I-V characteristics as in the present work but
with other materials and multiple barriers could be useful to estimate the interaction
times of electrons which are important in constructing solid state devices.
With the interaction time of tunneling particles being in general a question of
fundamental importance, the temperature dependent characteristics found here could
have implications in other physical situations too.

We would like to acknowledge technical support from C. Talero and L. Gomez in our experimental
setup and J.-G. Ram{\'i}rez for useful discussions. This work was funded by ``Programa Nacional de Ciencias
B{\'a}sicas" COLCIENCIAS (No. 120452128168), ``Convocatoria Programas 2012"
Vicerrector{\'i}a de Investigaciones and ``Proyecto Semilla" Facultad de Ciencias of
Universidad de los Andes (Bogot{\'a}, Colombia).


\begin{thebibliography}{99}
\bibitem{Note}
Note that the current density is proportional to $\Phi_{0}/s^{2}$ making it more sensitive to the s parameter variations rather than $\Phi_{0}$.
\bibitem{comment}
Note that this values are very similar to the ones obtained for other semiconductors in \cite {odonnell}.
\end{thebibliography}
\end{document}